\documentclass[aps,prl,twocolumn,showpacs,superscriptaddress,groupedaddress]{revtex4-1}
\usepackage{graphicx}
\usepackage{dcolumn}
\usepackage{bm}
\usepackage{longtable}
\usepackage{subfigure}
\usepackage{amsmath}
\usepackage{lineno}
\begin{document}
\def\Journal#1#2#3#4{{#1} {\bf #2}, #3 (#4)}

\def\NCA{Nuovo Cimento}
\def\NIM{Nucl. Instr. Meth.}
\def\NIMA{{Nucl. Instr. Meth.} A}
\def\NPB{{Nucl. Phys.} B}
\def\NPA{{Nucl. Phys.} A}
\def\PLB{{Phys. Lett.}  B}
\def\PRL{Phys. Rev. Lett.}
\def\PRC{{Phys. Rev.} C}
\def\PRD{{Phys. Rev.} D}
\def\ZPC{{Z. Phys.} C}
\def\JPG{{J. Phys.} G}
\def\CPC{Comput. Phys. Commun.}
\def\EPJ{{Eur. Phys. J.} C}

\topmargin 0pt
\title{Energy Dependence of $K\pi$, $p\pi$, and $Kp$ Fluctuations in Au+Au Collisions from $\rm \sqrt{s_{NN}}$ = 7.7 to 200 GeV }


\author{
L.~Adamczyk$^{1}$,
J.~K.~Adkins$^{21}$,
G.~Agakishiev$^{19}$,
M.~M.~Aggarwal$^{32}$,
Z.~Ahammed$^{49}$,
I.~Alekseev$^{17}$,
J.~Alford$^{20}$,
A.~Aparin$^{19}$,
D.~Arkhipkin$^{3}$,
E.~C.~Aschenauer$^{3}$
% ...
}

\address{$^{1}$AGH University of Science and Technology, Cracow 30-059, Poland}
% ...
\address{$^{3}$Brookhaven National Laboratory, Upton, New York 11973, USA}
% ...
\address{$^{17}$Alikhanov Institute for Theoretical and Experimental Physics, Moscow 117218, Russia}
% ...
\address{$^{19}$Joint Institute for Nuclear Research, Dubna, 141 980, Russia}
\address{$^{20}$Kent State University, Kent, Ohio 44242, USA}
\address{$^{21}$University of Kentucky, Lexington, Kentucky, 40506-0055, USA}
% ...
\address{$^{32}$Panjab University, Chandigarh 160014, India}
% ...
\address{$^{49}$Variable Energy Cyclotron Centre, Kolkata 700064, India}
% ...

\date{\today}

\begin{abstract}
A search for the quantum chromodynamics (QCD) critical point was performed by the STAR experiment at the Relativistic Heavy Ion Collider, using dynamical fluctuations of unlike particle pairs. Heavy-ion collisions were studied over a large range of collision energies with homogeneous acceptance and excellent particle identification, covering a significant range in the QCD phase diagram where a critical point may be located. Dynamical $K\pi$, $p\pi$, and $Kp$ fluctuations as measured by the STAR experiment in central 0-5\% Au+Au collisions from center-of-mass collision energies $\rm \sqrt{s_{NN}}$ = 7.7 to 200 GeV are presented. The observable $\rm \nu_{dyn}$ was used to quantify the magnitude of the dynamical fluctuations in event-by-event measurements of the $K\pi$, $p\pi$, and $Kp$ pairs. The energy dependences of these fluctuations from central 0-5\% Au+Au collisions all demonstrate a smooth evolution with collision energy.\\

\end{abstract}
\maketitle

\newpage
\setcounter{page}{1}

There are indications from some lattice quantum chromodynamics (QCD) calculations that at large values of baryon chemical potential, $\mu_{\rm B}$, the crossover between hadronic and partonic (quark-gluon) matter becomes a first order phase transition \cite{Lattice_CP}. If these lattice calculations are correct, there should be a critical point where the first order phase transition line ends. Enhanced fluctuations in final-state observables are one of the possible signatures of a phase transition, particularly if the phase transition occurs near a critical point. Critical opalescence is one example of critical behavior observed in classical systems \cite{CritOp1, CritOp2, CritOp3}. If there is a QCD critical point it is possible that similar enhanced fluctuations could be observed in measurements of particle multiplicities or net-charge \cite{Koch1,jeon, Stephanov_1999}.  
The moments of measured distributions are sensitive to the correlation length, $\xi$ \cite{Stephanov_2009}. A non-monotonic excitation function (observable as a function of energy) of the measured moments can indicate contributions from critical phenomena \cite{Stephanov_2011}. STAR has recently published the energy dependence of higher moments of the net-proton \cite{STARNetp} and net-charge \cite{STARNetq} distributions, which do not convincingly exhibit such behavior.
Dynamical relative particle number ($K\pi$, $p\pi$, and $Kp$) fluctuations are an observable that might also be sensitive to signals originating from the deconfinement phase transition \cite{GGZ04} or critical point \cite{ARS_2010}. These fluctuations provide a connection to globally conserved quantities including baryon number, strangeness, and charge, and approximately conserved quantities such as entropy \cite{GGM04}. In 2010-11 a search for the onset of partonic deconfinement and the QCD critical point was undertaken at the Relativistic Heavy Ion Collider (RHIC) at Brookhaven National Laboratory (BNL). This involved an ``energy scan'' of Au+Au collisions at the following beam energies in the two-nucleon center-of-mass system, $\rm \sqrt{s_{NN}}$: 7.7, 11.5, 19.6, 27, 39, 62.4, and 200 GeV. Dynamical fluctuations for all three of the aforementioned pairs of particle species were studied at each energy and are reported here.\\
\indent This data spans a wide range in beam energy, which corresponds to baryon chemical potentials from 24 to 421 MeV in central A+A collisions \cite{Cleymans}. This is the first time such measurements have been carried out over more than an order of magnitude in beam energy, with the same colliding species, and with the same detector at a collider facility. This allows for a suite of measurements to be performed at many energies, while minimizing corresponding changes in detector acceptance that is inherent to fixed target experiments. Experimental measurements at mid-rapidity avoid complications from the spectator region and can be directly compared to lattice QCD calculations and models based on the grand canonical ensemble \cite{Koch2006}.

The observable $\nu_{\rm dyn}$ was used to quantify the magnitude of the dynamical fluctuations \cite{nudyn1, nudyn2, nudyn3}. This observable reflects deviations of the particle number distributions from those of a statistical distribution (no interparticle or dynamical correlations) and was originally developed to study net-charge fluctuations. $\nu_{\rm dyn}$ provides a measurement of the dynamical variance of the difference between the relative number of two particle species \cite{nudyn1, nudyn2}. This takes the form of ($\frac{N_{A}}{<N_{A}>} - \frac{N_{B}}{<N_{B}>})^{2}$. The generalized definition of $\nu_{\rm dyn}$ is given by

\begin{align}
\nu_{\rm dyn,AB} = \frac{\left<N_{\rm A}(N_{\rm A}-1)\right>}{\left<N_{\rm A}\right>^{2}}
+ \frac{\left<N_{\rm B}(N_{\rm B}-1)\right>}{\left<N_{\rm B}\right>^{2}} \nonumber \\
- 2\frac{\left<N_{\rm A}N_{\rm B}\right>}{\left<N_{\rm A}\right>\left<N_{\rm B}\right>},
\label{nudyn}
\end{align}

\noindent where $N_{\rm A}$ and $N_{\rm B}$ are the numbers of particles of species $A$ and $B$ in a particular event, and the brackets denote their averages. The indices $A$ and $B$ can be replaced by $\pi$, $K$, or $p$ to construct the required form of $\nu_{\rm dyn}$. By definition, Eq. (\ref{nudyn}) is symmetric under the transposition of the particles $A$ and $B$. It is also independent of the detection efficiency in the region of phase space of interest here \cite{nudyn2}. If the underlying measured distribution has contributions only from uncorrelated particles, $\nu_{\rm dyn}$ will be exactly equal to zero. For non-statistical distributions, $\nu_{\rm dyn}$ can either be positive or negative, depending on which of the three terms dominate. Positive values of $\nu_{\rm dyn}$ are indicative of anti-correlations, while negative values of $\nu_{\rm dyn}$ reflect correlations. The dynamical component is thus measured relative to the statistical baseline ($\nu_{\rm dyn} = 0$).

A study of $K\pi$ fluctuations in Au+Au collisions at $\rm \sqrt{s_{NN}}$ = 200, 130, 62.4, and 19.6 GeV was previously carried out by the STAR experiment \cite{starkpiprl}. Measured dynamical $K\pi$ fluctuations were observed to be energy independent. A similar variable, called $\sigma_{\rm dyn}$, was studied by the NA49 collaboration \cite{NA49_kpi_fluc}. Significantly larger values of $\sigma_{\rm dyn}$ for $K\pi$ pairs were observed at beam energies near 6 GeV, which were interpreted by the NA49 collaboration as possibly due to enhanced fluctuations resulting from the onset of deconfinement. The current study takes advantage of a more than a factor of 10 increase in the number of recorded events available at several of the previously measured beam energies, several new beam energies, an addition of a Time of Flight (TOF) detector, a lower material budget at the center of the STAR detector, and an improved charged-particle reconstruction algorithm. 
These improvements have reduced the statistical and systematic uncertainties in the present results compared to those discussed in Ref. \cite{starkpiprl}.


The data presented here for $K\pi$, $p\pi$, and $Kp$ fluctuations were acquired by the STAR experiment \cite{STAR} at RHIC in minimum bias (MB) Au+Au collisions at $\rm \sqrt{s_{NN}}$ = 7.7, 11.5, 19.6, 27, 39, 62.4, and 200 GeV (3, 4, 15, 29, 10, 17, and 33 million events, respectively). The main particle tracking detector at STAR is the Time Projection Chamber (TPC) \cite{STARTPC}. The position of the collision vertex along the beam line was restricted to the center of the TPC to $\pm$ 30 cm at $\rm \sqrt{s_{NN}}$ = 19.6 to 200 GeV, and $\pm$ 50 cm at $\rm \sqrt{s_{NN}}$ = 7.7 and 11.5 GeV. A distance of closest approach (DCA) of a track to the primary event vertex of less than 1.0 cm was required to reduce the number of particles not originating from the primary collision vertex. Each track was required to have at least 15 fit points in the TPC, and a ratio of number of fit points to maximum possible number of fit points greater than 0.51. Collision centrality is determined (at all energies) using TPC charged particle tracks from the primary vertex in the pseudorapidity range $|\eta|< $ 0.5, with $p_{T} > 0.15$ GeV/$c$, and more than 10 fit points.

Particles were identified using a combination of the TPC and the recently completed TOF detector \cite{STARTOF, TPCTOF}. With these two detectors, the particle identification capabilities reach total momentum $p, \approx$ 1.8  GeV/$c$ for pions/kaons and $p \approx$ 3.0 GeV/$c$ for protons. The transverse momentum, $p_{T}$, range for pions and kaons was $p_{T} > 0.2$ GeV/$c$ and total momentum $p < $ 1.8 GeV/$c$, and for protons was $p_{T} > 0.4$ GeV/$c$ and $p < 3.0$ GeV/$c$.

Charged particle identification involved measured ionization energy loss, dE/dx, in the TPC gas and total momentum $p$ of the track. The energy loss of the identified particle was required to be less than two standard deviations, $\sigma$, from the predicted energy loss of that particle. Exclusion cuts were utilized to suppress misidentified particles. It was required that the measured energy loss of a pion(kaon) was more than 2$\sigma$ from the energy loss prediction of a kaon(pion). Similar exclusion cuts were used for proton identification. All charged particles in the interval $|\eta| < 1.0$ satisfying these cuts were included in the present analysis. Though the selected phase space for analysis overlaps with that for the centrality determination it was verified that $\nu_{dyn}$ was not affected by auto-correlations by performing a cross-check using separate regions of the TPC to calculate $\nu_{dyn}$ and determine centrality. The result for $\nu_{dyn}$ was consistent between these two methods.

Particle identification was also carried out by adding TOF information to that given by the TPC, which then provides a measurement of the mass-squared, $m^{2}$, for each track. Mass-squared cuts used for particle identification required an identified particle to be less than 2$\sigma$ from the predicted time-of-flight of that particle.
	
The final particle identification information uses both the TPC and TOF simultaneously, with the total acceptance for pions and kaons: $|\eta| < 1.0$, $p_{T} > 0.2$ GeV/$c$, and $p < $ 1.8 GeV/$c$, and for protons: $|\eta| < 1.0$, $p_{T} > 0.4$ GeV/$c$, and $p < $ 3.0 GeV/$c$. For particles with no TOF information, only the TPC dE/dx was used. 

The statistical error bars were obtained using a subsampling method and are generally small. The main sources of systematic errors in this study are from particle misidentification and electron contamination. These were estimated by relaxing the 2$\sigma$ TPC dE/dx cuts to 3$\sigma$, thereby increasing particle misidentification. This effect is most significant for $K\pi$ fluctuations and minimal for $p\pi$ fluctuations. For $K\pi$ fluctuations, the rate of kaon misidentification (integrated over all momenta) is as large as 17\% when using dE/dx alone. However, the pion contamination of the kaons is less than 4\% when using combined dE/dx+TOF information. Particle misidentification contributes a 3\% relative systematic error to $K\pi$ and $\approx$ 1\% to $p\pi$ and $Kp$ fluctuation measurements. Electron contamination provides an additional 5\% relative systematic uncertainty. Simulations based on the Ultra Relativistic Quantum Molecular Dynamics (UrQMD) model \cite{URQMD} indicated that the contributions to the present results from $p_{T}$-dependent inefficiencies are much smaller than the present uncertainties.


\begin{figure}[hbtp]
\centering
\includegraphics[width=0.505\textwidth]{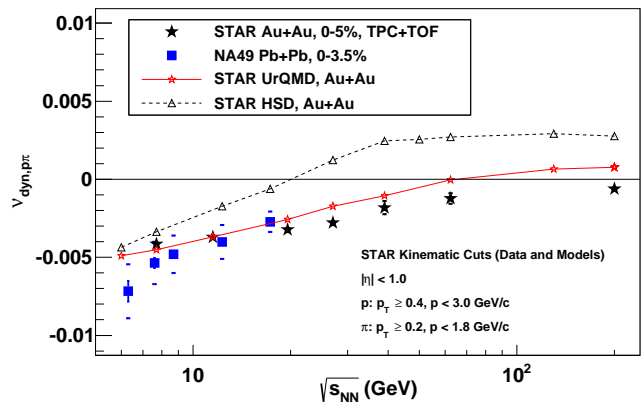}
\caption{$p\pi$ fluctuations as a function of collision energy, expressed as $\nu_{\rm dyn,p\pi}$. Shown are data from central (0-5\%) Au+Au collisions at energies from $\rm \sqrt{s_{NN}}$ = 7.7 to 200 GeV from the STAR experiment (black stars), predictions from UrQMD and HSD filtered through the same experimental acceptance (open stars and open triangles, respectively), and data from central (0-3.5\%) Pb+Pb collisions at energies from $\rm \sqrt{s_{NN}}$ = 6.3 to 17.3 GeV from the NA49 experiment (blue squares) \protect \cite{NA49_kpi_ppi}.}
\label{ppi}
\end{figure}

\begin{figure}[hbtp]
\centering
\includegraphics[width=0.50\textwidth]{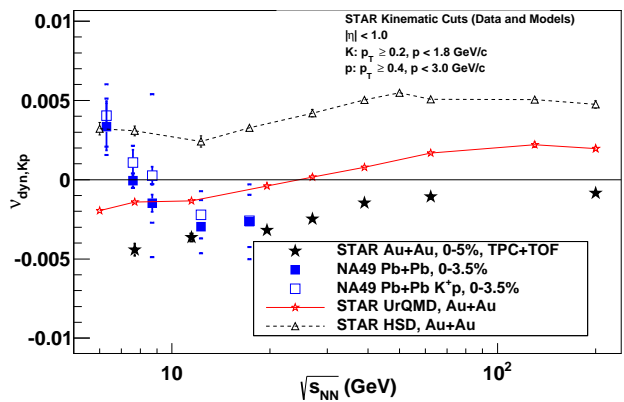}
\caption{$Kp$ fluctuations as a function of collision energy, expressed as $\rm \nu_{dyn,Kp}$. Shown are data from central (0-5\%) Au+Au collisions at energies from $\rm \sqrt{s_{NN}}$ = 7.7 to 200 GeV from the STAR experiment (black stars), predictions from UrQMD and HSD filtered through the same experimental acceptance (open stars and open triangles, respectively), and data from central (0-3.5\%) Pb+Pb collisions at energies from $\rm \sqrt{s_{NN}}$ = 6.3 to 17.3 GeV from the NA49 experiment (blue squares) along with charge-separated $K^{+}p$ fluctuations (open squares). \protect \cite{NA49_kp}.}
\label{kp}
\end{figure}

Dynamical $p\pi$ fluctuations in central 0-5\% Au+Au collisions as a function of the collision energy, $\rm \sqrt{s_{NN}}$, are shown in Fig. \ref{ppi}. Statistical error bars in all figures (where larger than the data point) are shown as the vertical lines and systematic errors are represented as caps above and below the data points. Figure \ref{ppi} shows that $\nu_{\rm dyn,p\pi}$ (stars) is negative across the entire energy range studied, is most negative at the lower energy Au+Au collisions, and becomes less negative as the energy is increased, eventually approaching zero at $\rm \sqrt{s_{NN}}$ = 200 GeV.  This indicates that protons and pions become less correlated as the collision energy is increased. 

The predominant source of correlated proton and pion production comes from the formation and decay of $\Delta$ resonances. Weak decays (such as from $\Lambda^{\rm 0}$ hyperon) are suppressed via the DCA cut described earlier. As the collision energy increases, the numbers of protons and anti-protons created via pair production also increases. Protons and anti-protons that are pair produced will not be correlated with the pions produced via, e.g., $\Delta$ resonances. Therefore if the rate of pair production exceeds the rate of resonance production, the relative correlation between protons and pions will decrease, leading to the observed energy dependence.

Also plotted in Fig. \ref{ppi} are two transport model predictions for the values of $\nu_{\rm dyn,p\pi}$ from the UrQMD (open stars) \cite{URQMD} and Hadron String Dynamics (HSD, open triangles) \cite{HSD} models, with the same kinematic acceptance cuts as the data. These transport models do not include a phase transition nor a critical point. UrQMD predicts negative values for $\nu_{\rm dyn,p\pi}$ at the lower energies, but positive dynamical $p\pi$ fluctuations above approximately $\rm \sqrt{s_{NN}}$ = 60 GeV. HSD predicts a similar qualitative trend, but crosses zero at approximately $\rm \sqrt{s_{NN}}$ = 20 GeV. In both models, the production rate of pair-produced protons and anti-protons grows with increasing energy, driving the prediction of $\nu_{\rm dyn,p\pi}$ positive at values that depend on the model. Both models are in relatively good agreement with the measured values of the present dynamical $p\pi$ fluctuations at $\rm \sqrt{s_{NN}}$ = 7.7 GeV, but HSD overpredicts at the other energies, while UrQMD overpredicts the present results above $\rm \sqrt{s_{NN}}$ = 19.6 GeV.

Figure \ref{ppi} also includes the dynamical $p\pi$ fluctuations as measured by the NA49 experiment, which used the observable $\sigma_{\rm dyn,p\pi}$ \cite{NA49_kpi_ppi}.
It is expressed as

\begin{align}
\rm \sigma_{dyn} = sgn(\sigma_{data}^{2}-\sigma_{mixed}^{2})\sqrt{|\sigma_{data}^{2}-\sigma_{mixed}^{2}|},
\label{sigmadyn}
\end{align}

\noindent where $\sigma$ is the relative width of the $K\pi$, $p\pi$, or $Kp$ distribution in either real data or mixed events.
The two variables are related as $\rm \sigma_{\rm dyn}^{2} \approx \nu_{\rm dyn}$ \cite{baym,nudyn_sigmadyn}. Because NA49 is a fixed target experiment, there are differences in kinematic acceptances at each beam energy and also between the two experiments. In the range $\rm \sqrt{s_{NN}}$ = 7.7 to 19.6 GeV, there is consistency between measurements of dynamical $p\pi$ fluctuations from both experiments.

Figure \ref{kp} shows dynamical $Kp$ fluctuations, measured with $\nu_{\rm dyn,Kp}$, as a function of the collision energy. The energy dependence observed in the most central (0-5\%) Au+Au collisions from $\rm \sqrt{s_{NN}}$ = 7.7 to 200 GeV (black stars) for $\nu_{\rm dyn,Kp}$ is similar to that observed for $\nu_{\rm dyn,p\pi}$ ({\it cf.} Fig. \ref{ppi}). The value of $\nu_{\rm dyn,Kp}$ is most negative at $\rm \sqrt{s_{NN}}$ = 7.7 GeV, becoming less negative and approaching zero as the energy is increased to $\rm \sqrt{s_{NN}}$ = 200 GeV, indicating a decreasing correlation between produced kaons and protons as the beam energy is increased. The UrQMD and HSD transport model predictions are also shown by the open stars and open squares, respectively. The UrQMD predictions for the dynamical $Kp$ fluctuations are similar to those for dynamical $p\pi$ fluctuations, which are negative at lower energies, then cross zero and become positive at higher energies. For $\nu_{\rm dyn,Kp}$, the HSD model prediction is always positive and almost energy-independent, unlike the prediction for $\nu_{\rm dyn,p\pi}$, which was qualitatively similar to the UrQMD prediction. One difference between the two models is that they treat resonance decays in different ways, so the final state correlations are model dependent \cite{HSD_Resonance}.

Figure \ref{kp} also includes the measured dynamical $Kp$ fluctuations from $\sigma_{\rm dyn,Kp}$ converted to $\nu_{\rm dyn,Kp}$ from the NA49 experiment \cite{NA49_kp}, which used similar central (0-3.5\%) Pb+Pb collisions. Unlike the energy dependence that is presently observed below energies of $\rm \sqrt{s_{NN}}$ = 11.5 GeV, the NA49 results for dynamical $Kp$ fluctuations trend toward zero (and ultimately cross zero), and become positive below $\rm \sqrt{s_{NN}}$ = 7.6 GeV. The energy and charge dependence of $\nu_{\rm dyn,Kp}$ in central (0-5\%) Au+Au collisions have negative values and do not cross zero \cite{sqm2011_kp_ppi}. Therefore, the change in sign of the inclusive $Kp$ dynamical fluctuations is not reproduced.
The different momentum space coverage between NA49 (forward rapidity, $p >$ 3.0 GeV/$c$) and STAR (mid-rapidity) and its effects on $\nu_{\rm dyn}$ was discussed in Ref. \cite{NA49_nudyn}. They find that $\nu_{dyn}$ at low SPS energies (20A and 30A GeV) has a dependence on the phase space coverage that explains the differences between NA49 and STAR results. However, it was demonstrated that $\nu_{dyn}$ depends on experimental azimuthal acceptance \cite{HaslumThesis}. Limited experimental acceptance impacts the detection of particle pairs from resonance decay depending on whether the decay daughters are emitted in the same direction, back-to-back, or not correlated and consequently changes the measured value of the fluctuations. A detector with full 2$\pi$ azimuthal acceptance will not observe a difference in $\nu_{dyn}$ regardless of the direction of emitted decay daughter pairs.

\begin{figure}[]
\centering
\includegraphics[width=0.50\textwidth]{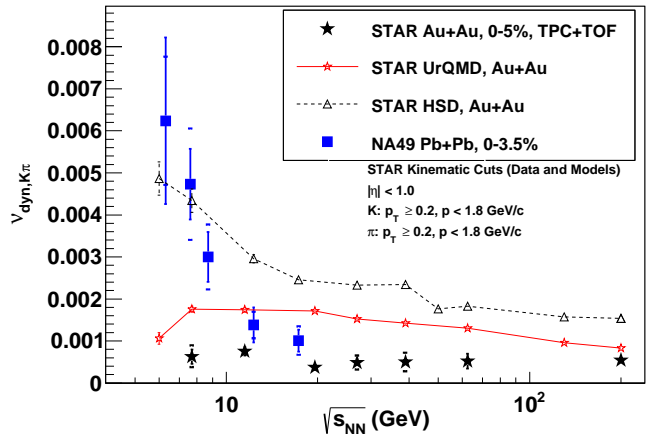}
\caption{$K\pi$ fluctuations as a function of collision energy, expressed as $\nu_{\rm dyn,K\pi}$. Shown are data from central (0-5\%) Au+Au collisions at energies from $\rm \sqrt{s_{NN}}$ = 7.7 to 200 GeV from the STAR experiment (black stars), predictions from UrQMD and HSD filtered through the same experimental acceptance (open stars and open triangles, respectively), and data from central (0-3.5\%) Pb+Pb collisions at energies from $\rm \sqrt{s_{NN}}$ = 6.3 to 17.3 GeV from the NA49 experiment (blue squares) \protect \cite{NA49_kpi_ppi}.}
\label{kpi}
\end{figure}

Figure \ref{kpi} depicts the values of dynamical $K\pi$ fluctuations in central (0-5\%) Au+Au collisions from $\rm \sqrt{s_{NN}}$ = 7.7 to 200 GeV, as measured by $\nu_{\rm dyn,K\pi}$ from the STAR experiment (black stars). Unlike $\nu_{\rm dyn,p\pi}$ and $\nu_{\rm dyn,Kp}$, the inclusive charged particle $\nu_{\rm dyn,K\pi}$ are always positive.
This indicates that for produced kaons and pions either the variance of the two particle species dominates and/or there is an anti-correlation ($\left<N_{\rm K}N_{\pi}\right> < 0$) between the produced particles. The primary resonances that contribute to $\nu_{\rm dyn,K\pi}$ are the $\rm K^{*}(892)$ and $\phi(1020)$.
A study of the resonance contribution to $K\pi$ fluctuations using UrQMD was shown in \cite{KresanFriese2006}. The measured STAR experimental value of $\nu_{\rm dyn,K\pi}$ is approximately independent of collision energy in central (0-5\%) Au+Au collisions from $\rm \sqrt{s_{NN}}$ = 7.7 to 200 GeV. 

The predictions for $\nu_{\rm dyn,K\pi}$ from UrQMD (open stars) and HSD (open triangles) tend to overpredict the magnitude of the fluctuations at high energies, but approximate the qualitative trend from the observations. UrQMD is also consistent with a flat trend. HSD predicts increased fluctuations at the lower energies and agrees with the measurements of $\sigma_{\rm dyn,K\pi}$ by the NA49 experiment (blue squares) \cite{NA49_kpi_ppi}. The differences between the two predictions are primarily the result of the treatment of resonance decay. 
The measured energy dependence of the dynamical $K\pi$ fluctuations in central (0-5\%) Au+Au and central (0-3.5\%) Pb+Pb collisions is similar between $\rm \sqrt{s_{NN}}$ = 11.5 and 19.6 GeV. Below $\rm \sqrt{s_{NN}}$ = 11.5 GeV, there is a large difference between measurements from the two experiments, with the STAR results remaining approximately energy independent and those from NA49 increasing rapidly. Table \ref{MultTable} shows the STAR efficiency uncorrected identified particle numbers used in this analysis, while the NA49 values can be found at Ref.~\cite{NA49_kp}.

Examining the energy dependence of the dynamical $K\pi$, $p\pi$, and $Kp$ fluctuations in the central (0-5\%) Au+Au collisions from $\rm \sqrt{s_{NN}}$ = 7.7 to 200 GeV, there do not appear to be any trends in the beam-energy dependence that represent clear evidence of critical fluctuations or the deconfinement phase transition. The two primary interpretations are that dynamical particle number fluctuations may not be sensitive to these phenomena, or that the phase transition at these baryon chemical potentials does not cross through a critical point.

\begin{table}
\begin{center}
  \begin{tabular}{ |c || c | c | c |  }
    \hline
     $\sqrt{s_{\rm NN}}$  (GeV) & $<K^{\pm}>$&  $<\pi^{\pm}>$ & $<p^{\pm}>$ \\ \hline
     200  &34$\pm2$ & 366$\pm$18 & 23$\pm1$ \\ \hline
     62.4 &35$\pm$2 & 345$\pm$17 & 25$\pm$1 \\ \hline
     39     &31$\pm$2 & 317$\pm$16 & 23$\pm$1 \\ \hline
     27     &24$\pm$1 & 282$\pm$14 & 22$\pm$1 \\ \hline
     19.6  &25$\pm$1 & 272$\pm$14 & 26$\pm$1 \\ \hline
     11.5  &21$\pm$1 & 209$\pm$11 & 32$\pm$2 \\ \hline
     7.7     &16$\pm$1 & 161$\pm$8 & 39$\pm$2 \\
    
    \hline

  \end{tabular}
\caption{Average number of efficiency uncorrected identified particles measured by STAR and used in the analysis of $\nu_{\rm dyn}$ (0-5\% centrality only).}
\label{MultTable} 
\end{center}
\end{table}

In summary, STAR has made measurements of the dynamical $K\pi$, $p\pi$, and $Kp$ fluctuations in Au+Au collisions across a broad range in collision energy from $\rm \sqrt{s_{NN}}$ = 7.7 to 200 GeV. This is the first time these measurements have been carried out over more than an order of magnitude in the collision energy, with the same colliding species, and with the same detector at a collider facility.
The dynamical $p\pi$ and $Kp$ fluctuations (measured with $\nu_{\rm dyn}$) in central (0-5\%) Au+Au collisions from $\rm \sqrt{s_{NN}}$ = 7.7 to 200 GeV are negative and approach zero as the collision energy increases, indicating less correlation between the measured particles. The dynamical $K\pi$ fluctuations in central (0-5\%) Au+Au collisions from $\rm \sqrt{s_{NN}}$ = 7.7 to 200 GeV are always positive and approximately independent of the collision energy. The beam-energy dependence for dynamical fluctuations of the three pairs of particles evolve smoothly with collision energy in central (0-5\%) Au+Au collisions and do not exhibit any significant non-monotonicity that might indicate the existence of a phase transition or a critical point in the QCD phase diagram. The study of the multiplicity scaling of the energy dependence of the particle number fluctuations and the charge dependence of these results may provide additional insight into the mechanisms that cause the observed fluctuations and correlations.\\

We thank the RHIC Operations Group and RCF at BNL, the NERSC Center at LBNL, the KISTI Center in Korea, and the Open Science Grid consortium for providing resources and support. This work was supported in part by the Office of Nuclear Physics within the U.S. DOE Office of Science, the U.S. NSF, CNRS/IN2P3, FAPESP CNPq of Brazil,  the Ministry of Education and Science of the Russian Federation, the NNSFC, the MoST of China (973 Program No. 2014CB845400), CAS, the MoE of China, the Korean Research Foundation, GA and MSMT of the Czech Republic, FIAS of Germany, DAE, DST, and CSIR of India, the National Science Centre of Poland, National Research Foundation (NRF-2012004024), the Ministry of Science, Education and Sports of the Republic of Croatia, and RosAtom of Russia.



\bibliography{all}
\bibliographystyle{apsrev4-1}

\end{document}